\documentclass[lettersize,journal]{IEEEtran}
\usepackage{cite}
\usepackage{amsmath,amssymb,amsfonts}
\usepackage{algorithmic}
\usepackage{graphicx}
\usepackage{textcomp}
\usepackage[caption=false,font=normalsize,labelfont=sf,textfont=sf]{subfig}
\usepackage{stfloats}
\usepackage{comment}
\usepackage{verbatim}
\usepackage{epstopdf}
\usepackage{tabularx}
\usepackage{url}
\def\BibTeX{{\rm B\kern-.05em{\sc i\kern-.025em b}\kern-.08em
    T\kern-.1667em\lower.7ex\hbox{E}\kern-.125emX}}
\usepackage{balance}
\begin{document}
\title{Ultralow-power standoff acoustic leak detection}
\author{Michael P. Hasselbeck
\thanks{Submit date 31 October 2025}
\thanks{M.P. Hasselbeck is with MicroPhonon LLC \\
www.microphonon.com \\
E-mail: mph@microphonon.com}}

\maketitle

\begin{abstract}
 An automated, standoff acoustic leak detection scheme has been designed, built, and tested. It merges the principles of glass breakage and smoke detection to alert for the presence of leaks emanating from pressurized plumbing. A simulated water leak flowing at 0.15 l/min has been reliably detected at a standoff distance of more than 10 m. The device is also effective at identifying the presence of leaks located behind surfaces such as walls, doors, floors, and ceilings.  The anticipated application is as an autonomous, battery-powered, remote wireless node. All signal processing and analysis takes place on the edge with no need to stream audio data to the cloud. Sensor status is conveyed on-demand with only a few bytes of information, requiring minimal bandwidth. Power consumption is the range of 20--200 $\mu$W, depending on the amount of environmental noise and desired sensor latency. To attain optimum sensitivity and reliability, the hardware operates at acoustic frequencies well above the range of human conversations, making eavesdropping impossible.
Development has been done with water escaping from pressurized plumbing, but the sensor concept can be used effectively to detect gas leaks. 
\end{abstract}


\section{Introduction}
There are two methods currently available for autonomous water leak detection: contact moisture sensors \cite{1} and smart flow meters \cite{2}. They work well and can provide protection in many applications. Identifying the onset of very small leaks, however, can be difficult especially in situations where installing existing products may not be easy or convenient. This paper describes an automated, inexpensive, battery-powered, smart acoustic leak detector concept that addresses these problems and can augment the capabilities of a building monitoring system. 

When pressurized water escapes from a seam, crack, or loose fitting on plumbing, broadband high
frequency acoustics are emitted. These signals can travel an extended distance ($>$ 10 m) in free-space
and be detected by a remote sensor \cite{3,4}. This suggests a standoff method for water leak detection that is
substantially different compared to conventional contact moisture sensors and various types of smart
flow meters installed on or inside of plumbing lines.

Using acoustics to detect leaks is a well established approach and a variety of commercial devices are available. These sensors are generally sophisticated, expensive, and require a trained technician to operate \cite{5,6}. It is desirable to completely automate the leak detection process with a smart device that can be deployed as a remote sensor node on a wireless network.

The acoustic leak sensor combines the principles of glass breakage and smoke detectors. Glass breakage occurs on a timescale of a few seconds or less, requiring continuous data acquisition at rates approaching 1
kHz or more \cite{7}. Almost all pressurized plumbing leaks are persistent, which allows for sub-Hz sampling rates and
exceptional battery life. In a typical application, two AAA alkaline batteries should power the
device for multiple years.

Glass breakage has the advantage of being generally louder than the ambient environment so that
detector sensitivity is not an issue. Small water leaks, especially when separated from the sensor
by acoustic barriers such as walls, are likely to be at or below the environmental background. Much
higher sensitivity and noise mitigation are crucial. This can be addressed by active (analog and
digital) and passive (mechanical) filtering along with judicious use of statistical analysis.

The acoustic approach to water leak detection described here is not capable of
identifying all water issues. It will not respond to leaking roofs or melting ice, for example. The nature
of the pressurized orifice is also important. A pinhole leak will produce a Laminar
flow jet that emits a much weaker acoustic signal -- perhaps undetectable -- compared to a jagged crack
or loose fitting at the same backing pressure. 

Acoustic sensing can be used to complement standard
leak detection techniques and will be especially beneficial when monitoring large areas. Anticipated use cases are in smart homes and buildings, security systems, irrigation, dental and medical offices, critical plumbing installations, insurance loss mitigation, and water conservation. Although the design been developed for water leaks, it is expected to be useful in other applications that involve pressurized gases such as air, refrigerants, or steam.

It is important to emphasize that there is no attempt to make quantitative measurements of flow. The sensor is listening for a persistent acoustic signal that indicates the possible presence of a leak. The leak rate is irrelevant; we are only interested in whether or not one is present. It targets the smallest leaks, below flow rates where utility meters may not respond.

The acoustic leak detector is agnostic to the host hardware and will work with any controller having an I$^2$C interface and
3V3 power bus.
Best performance is attained in an environment with a stable, relatively quiet acoustic
background. Human activity will often produce noise that will impair the sensor’s ability to hear leaks.
The formation of damp spots, puddles, and similar indicators of water escaping, however, are usually obvious and
quickly noticed. Highest reliability for an acoustic sensor will occur in conditions where people are not
present.

\section{Design}
\begin{figure}[!t]
\centerline{\includegraphics[width=3in]{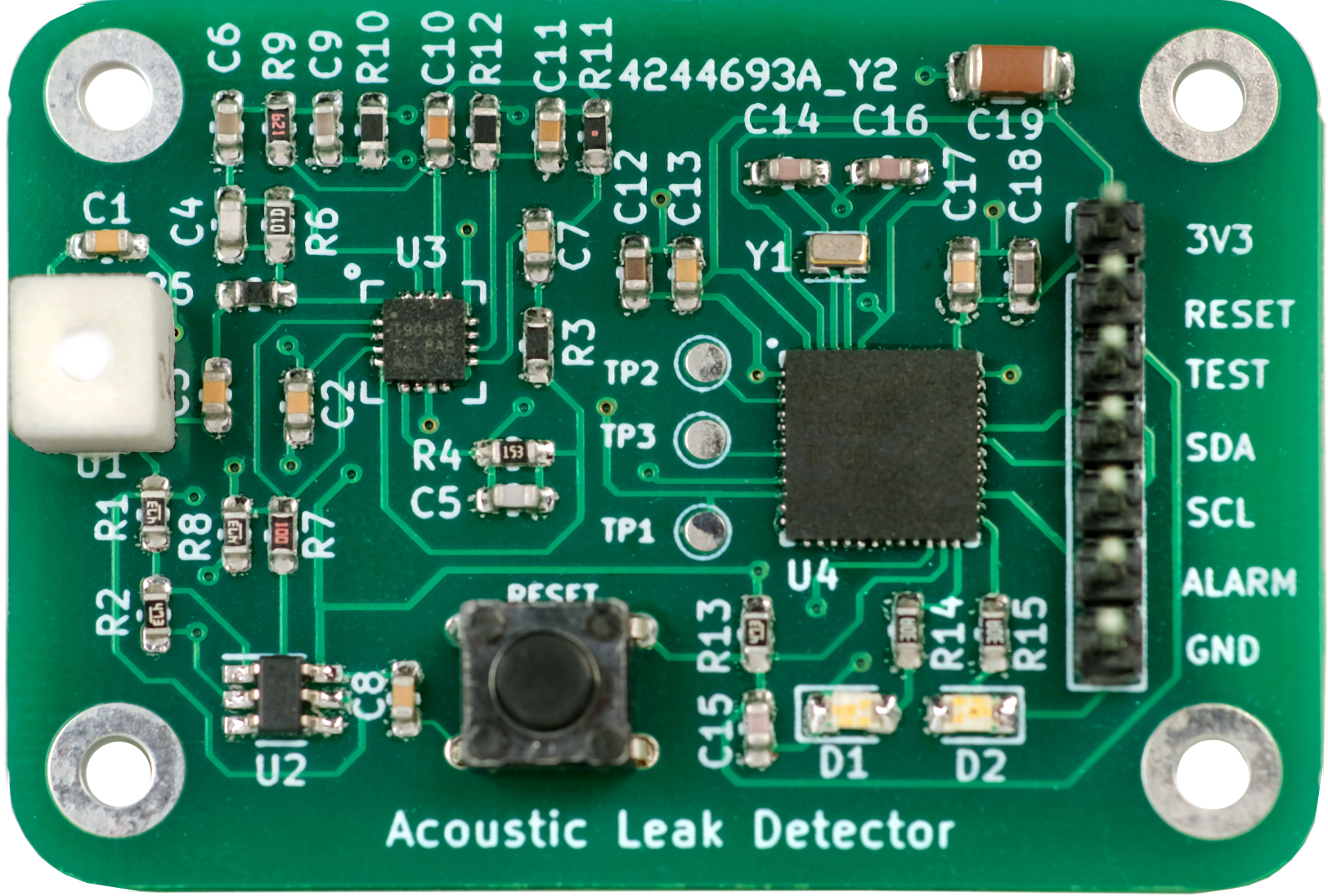}}
\caption{The sensor PCB measures 48 mm x 33 mm. A surface mount MEMS microphone is enclosed by a rectangular chamber to form a Helmholtz resonator (far left). The 7-pin connection header is used for programming the MCU, I$^2$C communication, and 3V3 dc power.}
\label{fig:board}
\end{figure}

The sensor is fabricated on a double-sided PCB using low cost, commercial electronic components (see Fig.~\ref{fig:board}). It is
designed to work with a host controller as part of a low power wireless system with a well-filtered 1.8 to 3.6 Vdc power source.  All sensor training, signal
processing, and analysis occurs on the device; no audio is ever sent to the cloud. Two-way controller-sensor communication takes places via I$^2$C. Firmware, a system flow chart, and design files are available in the GitHub repository \cite{8}. 

\subsection{Frequency Filtering}
\label{sec:filtering}
It is essential to separate weak leak signals from the ambient background to prevent false triggers and reduce the likelihood of false alarms. Leaks are sufficiently broadband that they produce discernible acoustic energy at frequencies above 6 kHz; these are the frequencies targeted by the sensor. The lower frequencies are attenuated by a combination of mechanical, analog electronic, and digital filtering.

\begin{figure}[!t]
\centerline{\includegraphics[width=2in]{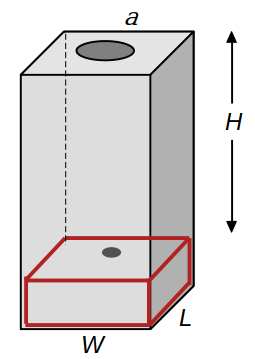}}
\caption{Design diagram of the Helmholtz resonator. A partially open rectangular chamber is placed over a MEMS microphone (red lines). The resonator wall thickness $t$ is not shown for clarity.}
\label{fig:resonator}
\end{figure}

Mechanical filtering is implemented with a miniature Helmholtz resonator that is formed by a rectangular plastic chamber glued above a surface-mount MEMS microphone (Knowles SPU0410). Acoustic energy is admitted through a circular hole of radius $a$. Referring to Fig.~\ref{fig:resonator}, the resonator center frequency $f_0$ is given by \cite{9}:

\begin{equation}f_0 = \frac{va}{2}\sqrt{\frac{1}{\pi V t'}}\label{eq:1}\end{equation}

\noindent where $v$ = 343 m/s is the speed of sound and  $V = LWH$ is the partially enclosed air volume above the microphone surface.
The rectangular footprint of the microphone is $L$ = 3.9 mm and $W$ = 3.2 mm; $H$ is the interior height of the chamber. The wall thickness $t$ affects the acoustic response of the hole via the Rayleigh end correction \cite{9}:

\begin{equation}t' = t + 1.7a.\label{eq:2}\end{equation}

\noindent The $Q$ of the resonator is estimated as:

\begin{equation}Q \approx 2~\sqrt{\frac{V}{\pi}\left(\frac{t'}{a^2}\right)^3}.\label{eq:3}\end{equation}

\noindent Setting $H$ = 3.5 mm and $a$ = $t$ = 1 mm gives an estimate of $f_0$ = 8.9 kHz and $Q$ = 33. The dimensions are not critical. 

The resonator can be inexpensively 3-D printed and is visible on the far-left side of the PCB in Fig.~\ref{fig:board}. In addition to frequency filtering, testing indicates the Helmholtz resonator approximately doubles sensitivity as measured by the standoff detection distance.

The microphone signal is amplified 40 dB and sent to a 2-stage, 4th-order Sallen-Key high-pass filter with 3 dB cutoff at $\sim$8 kHz. This is followed by a 23 dB post-filter gain stage. Natural roll-off at high frequencies makes band-pass filtering unnecessary; all filter poles are placed on the low frequency side of the cutoff frequency to sharpen the response. To simplify the design, the analog front-end uses a virtual ground at approximately half the supply voltage.

The electronically filtered signal is sampled by the 12-bit ADC in the MCU (Texas Instruments MSP430FR5994). A set of 256 data points are collected at 33.3 kS/s, corresponding to a 7.68 ms duration temporal acquisition. The data is checked for ADC overload and tagged as noisy if necessary. If no overload has occurred, the MCU transfers the data via Direct Memory Access to the Low Energy Accelerator (LEA) which performs an FFT of the time data using a uniform window, as appropriate for broadband white noise. The LEA calculates the FFT $>$ 10$\times$ faster and more
efficiently than an ARM-Cortex M0+. The spectral data is rendered as an array of complex numbers separated by 130.2 Hz frequency steps. A 34 point subset in the range 7 to 11.5 kHz is selected. The spectral energy at each frequency step is calculated and then summed to provide the total acoustic energy in the frequency window. This sum represents one point in an accumulated data set that is analyzed statistically by the MCU.

The combination of mechanical, analog, and digital filtering greatly
reduces the influence of background noise, dramatically reduces the potential of ADC overload, and
eliminates the possibility of intelligible conversations being discerned if the sensor network becomes
compromised, i.e.~eavesdropping is impossible.

\subsection{Sensor Training}

Sensitivity is set automatically at startup and requires the device to learn about the ambient background during a short training session.  The acoustic environment should be nominally quiet and leak-free. Frequency-filtered audio streams of duration 7.68 ms (see Section \ref{sec:filtering})
are gathered at a 1 Hz rate and analyzed for their spectral energy content. The size of the training set is user-configurable in the range 10 to 255; e.g.~30 samples are acquired in 30 seconds.  Analysis takes place entirely in-situ: no audio is ever sent to the cloud.  

A training session establishes a baseline signal level characterized by an average $(\bar{x}_B)$ and standard deviation $(\sigma_B)$. If statistical analysis of this data
determines that the background acoustic level is sufficiently stable $(2\sigma_B < \bar{x}_B)$, the
MCU firmware transitions to monitor mode. If not, the background is too noisy for reliable operation and a new training session takes place. If the sensor cannot find a stable background, it will loop
indefinitely in the training mode.

To obtain additional noise rejection and reliability, the acoustic data stream is checked for ADC overload.
Excessively loud signals from intermittent impulse noise are discarded and the individual acquisition is repeated. This continues until
the background acoustic environment is sufficiently quiet and stable. 

\subsection{Polling}
When monitoring, maximum battery life is attained when the sensor spends extended time in its low-power sleep mode. At timed intervals, the MCU wakes up, makes an acoustic measurement, records the analysis result, then goes back to sleep. The user-selected polling period ($\tau$ = 1 to 30 seconds) is sourced to a low-power clock that is referenced to a board-mounted 32768 Hz crystal. Longer polling periods consume less battery energy at the expense of sensor latency.

The microphone is pulse-biased and the amplifier and filter stages (Texas Instruments TLV9064S) are gated on-off to increase battery life. To
illustrate, if a single audio acquisition occurs every 2 seconds, the sensor will be in sleep
mode $>$ 99 percent of the time.  Total current draw in sleep mode is $<$ 4 $\mu$A
corresponding to $\sim$12 $\mu$W of continuous power drain. Each acoustic sampling event consumes $\sim$140
$\mu$J with a momentary peak current of about 3.2 mA.

The sensor acquires a constantly updated time-series data set. Because acoustic environments can vary widely depending on the application, the user must specify both the event data count ($N$ = 10 to 255) and polling period $(\tau)$. The product $N\tau$ is the duration of the event acquisition window. This window slides in time: the oldest event is discarded as each new event is added.

Configuring the event acquisition window requires some situational awareness. A robust, reliable deployment should anticipate the possible appearance of nominal, sustained background noise in the sensor frequency passband. This noise could be incorrectly interpreted as a leak. Examples are open faucets, running shower heads and dishwashers, and audio entertainment equipment.  To avoid triggering an alarm, $N\tau$ should be set longer than any anticipated persistent, high-frequency background sources. In an occupied residence, 20 minutes or more may be appropriate. The window could be as short as 2 minutes if the area being monitored is expected to be unoccupied and quiet, e.g.~an attic or crawl space.   

\subsection{Statistical Analysis}

Each polling cycle analyzes the acoustic signal and designates it as one of the following
three events: \textbf{quiet}, \textbf{leak}, or \textbf{noise}. These events are accumulated in three separate data arrays. The
value 1 is entered if that specific event occurs; otherwise it is counted as 0. For example: if a leak
signal is detected, a 1 is added to the \textbf{leak} array while 0 is placed in the \textbf{quiet} and \textbf{noise} arrays. As new
events are added, the oldest events are removed, i.e. the fixed-size data arrays slide in time.
All data arrays are the same size ($N$) and contain elements that are either 0 or 1. The training session sets the sensitivity level for detecting a leak signal, which is one standard
deviation above the noise floor: $\bar{x}_B + \sigma_B$.

The following takes place on each polling cycle: The filtered audio entering the ADC is first checked for saturation. If an
excessively loud signal is present, a \textbf{noise} event is recorded, and the
firmware enters a sleep state to wait for the next cycle.

 If there is no ADC overload, the spectral energy ($x_0$) derived
from an FFT of the temporal data is checked. If $x_0 < \bar{x}_B + \sigma_B$, no leak is detected and a \textbf{quiet} event is
recorded. If $x_0 > \bar{x}_B + \sigma_B$, additional processing commences: four sequential
acoustic energy measurements are made; these are separated by 45 ms corresponding to a sampling rate of
about 22 Hz. If any of these additional acquisitions saturate the ADC, further processing aborts, an
environmental \textbf{noise} event is recorded, and the polling loop pauses for the
next iteration.

If all five filtered audio signals are within the dynamic range of the ADC, their individual spectral
energies ($x_i$) are used to calculate the average energy: 

\begin{equation}\bar{x} = \frac{1}{5}\sum_{i=0}^{4}x_i\label{eq:4}\end{equation}

\noindent and standard deviation $(\sigma)$. If $2\sigma > \bar{x}$,
excessive signal fluctuations are present. This indicates a level of environmental noise that
prevents reliable leak detection. This polling cycle is recorded as a \textbf{noise} event and the program waits in low-power mode for the next timed iteration.

If the acoustic energy fluctuations are sufficiently small ($2\sigma < \bar{x}$), the next step is to check if all five spectral
energies $(x_i)$ exceed the leak signal threshold: $x_i > \bar{x}_B + \sigma_B$. When this occurs, a \textbf{leak} event is recorded. Otherwise, this loop iteration is associated with impulse noise, but
recorded as a \textbf{quiet} event. 

Performing the above sequence of tests and checks ensures that only persistent, stable signals in the
target acoustic frequency range are counted as \textbf{leak} events. This is a design trade-off between
sensitivity and reliability that is needed to reduce false alarms. 

We define the individual sums of the \textbf{quiet}, \textbf{leak}, and \textbf{noise} arrays as $Q$, $S$, and $R$, respectively. If $S$ become an appreciable portion of the event set size $(N)$, an alarm condition may
exist. Ideally, all events in the window will be identified as leaks, i.e~when $S = N$. In practice, environmental noise,
irregularities in water flow, pressure fluctuations, the presence of bubbles, or other acoustic
disturbances may be recorded as quiet or noise events -- not leaks. To account for non-ideal signals, the
alarm trigger threshold $(T)$ can be independently set below the event set size $(T < N)$. This threshold is not critical;
testing has shown that setting $T$ to an integer value that is 80 to 90 percent of $N$ is a good choice. 

A potential leak is identified when $S \geq T$; the MCU provides an immediate alert by switching a digital
alarm line to the high state (Logic 1). The host controller can monitor this line with an interrupt. An alarm is also conveyed by setting a status byte that the controller can retrieve on the I$^2$C serial interface.
The real-time counts $Q$, $S$, and $R$ are also available on demand. Recall that every polling cycle writes 1 to one array and 0 to the other two so that $Q + S + R = N$.

An essential design premise is that the acoustics produced by small leaks will be steady and persistent. This is not anticipated to be the situation with large leaks, which will likely be loud and erratic.

The sensor recognizes when it is operating in a noisy environment. This occurs when frequent, fluctuating, high-frequency acoustics emerge at levels significantly higher than were present during the training session. Noise will impair the sensor's ability to hear weak leak signals and may provide an indication of anomalous activity in the vicinity. The noise can arise from many causes such as water escaping at a high flow (i.e.~large leaks) or even unauthorized human activity. The latter might occur during an unauthorized break-in to what should be quiet, unoccupied property. When the two conditions $S < T$ and $S + R \geq T$ exist, then the sensor is hearing significant noise. The noise status byte is set and the controller is alerted with an interrupt.

Sensor status and array counts are available to the controller by issuing commands using the I$^2$C serial interface. The monitoring parameters are also configured via the I$^2$C bus using a simple command set \cite{8}. 

\section{Performance Testing}

Establishing sensor performance benchmarks is challenging because of varying background noise, the nature of the leak orifice, and environmental conditions. The latter may be dynamic and highly nonlinear. For example, a high concentration of water droplets around a leak can alter the local humidity, which in turn affects sound attenuation at different frequencies and at different points on the propagation path \cite{3}. To gain insight on performance and potential applications, a  series of experiments are described that characterize sensitivity and standoff detection with and without acoustic obstacles present.

\subsection{Free-Space Standoff Detection}
\label{sec:freespace}
Water leaks are simulated using a garden irrigation mister (Home Depot 7760F) at residential utility pressure. Flow is measured as 0.15 l/min, emitted in a 90 degree spray pattern. Reliable detection is attained at free-space standoff distances exceeding 10 m, both indoors and outdoors. 

\begin{figure}[!t]
\centerline{\includegraphics[width=3.5in]{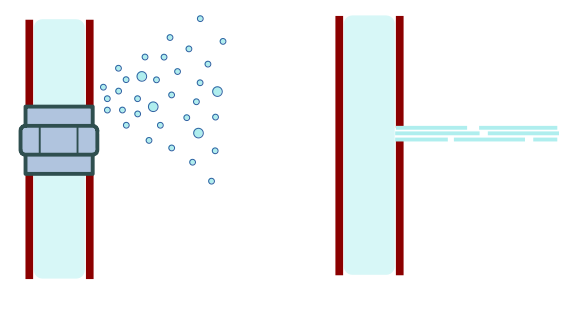}}
\caption{Two different plumbing leaks. Left: A dispersed spray appears at a loose fitting or crack. Right: A pinhole leak produces a Laminar jet. The leak on the left is much easier to detect because of its stronger high-frequency acoustics. }
\label{fig:leaks}
\end{figure}

Standoff detection changes dramatically depending on the nature of the leak. Referring to Fig.~\ref{fig:leaks}, the sketch on the left represents a spray pattern that might be anticipated at a loose pipe connection.  This leak is accompanied by strong high-frequency acoustics. The drawing on the right depicts a Laminar flow jet resulting from a pinhole in the pipe.

A 0.38 mm diameter hole drilled into a copper pipe produces a collimated water jet flowing at 0.23 l/min. This is $\sim$50 percent greater flow as compared to the mister, but with much weaker high-frequency content. This causes the standoff detection distance to drop by more than 3$\times$. Doubling the hole diameter increases the flow to more than 0.7 l/min, but the standoff detection distance is still well below that obtained with a spray.

This is clear evidence that the leak rate is not well correlated with the detected sound level. Since most plumbing leaks occur at cracks, seams, or loose fittings, they will likely produce a spray and concomitant high-frequency acoustics. These are the frequencies monitored by the sensor.

\subsection{Leaks Behind Walls}

When an acoustic barrier such as a wall is in the path between a leak and the sensor, sensitivity is greatly reduced. This is quantified in measurements with the irrigation mister (Section \ref{sec:freespace}) located $\sim$15 cm below boards made from various wall materials. Care is taken to ensure that sound must propagate through the material, not around it. Results are summarized in Table \ref{tab1}. Although the standoff detection distance is significantly reduced compared to a free-space path, the sensor is still able to reliably detect the simulated leak when it is positioned sufficiently close to the wall surface. Note the expected increased attenuation with thicker material sections.

\begin{table}
\centering
\caption{Acoustic Attenuation of Wall Materials}
\setlength{\tabcolsep}{3pt}
\begingroup
\renewcommand{\arraystretch}{1.5}
\begin{tabular}{|p{100pt}|p{110pt}|}
\hline
Material & 
Detection distance from surface  \\
\hline
1.3 cm thick gypsum board$^{\mathrm{a}}$ & 56-61 cm \\
1.3 cm thick gypsum board$^{\mathrm{a}}$ + 2.5 cm fiberglass insulation & 43-46 cm \\
0.6 cm thick plywood$^{\mathrm{b}}$ & 71-76 cm \\
1.3 cm thick plywood$^{\mathrm{b}}$ & 43-46 cm \\

\hline

\multicolumn{2}{p{200pt}}{$^{\mathrm{a}}$Also known as drywall or sheetrock.}\\
\multicolumn{2}{p{200pt}}{$^{\mathrm{b}}$Also known as plyboard or kryssfaner.}
\end{tabular}
\endgroup
\label{tab1}
\end{table}

\begin{figure}[!t]
\centerline{\includegraphics[width=3.5in]{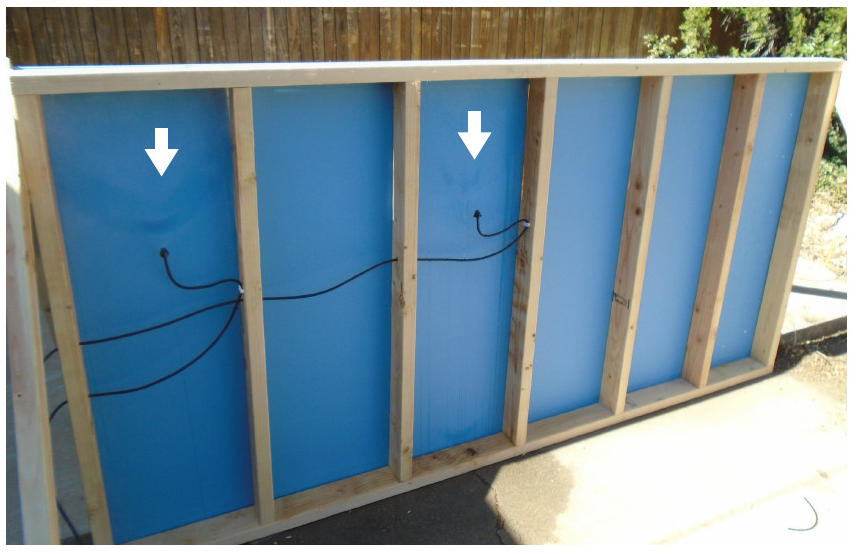}}
\caption{Photograph of partially completed test structure used to evaluate acoustic detection of water leaks behind gypsum board. Arrows show the fixed locations of two independently activated simulated leaks. A second gypsum board (not shown) seals the structure for testing.}
\label{fig:wall}
\end{figure}

Leak detection behind walls is further characterized with the test structure displayed in Fig.~\ref{fig:wall}. It measures 2.44 m x 1.22 m with an interior depth of 10 cm. Six internal sections are defined by 10 cm x 5 cm cross-section wooden vertical studs separated by 40 cm. This represents standard  wall construction in the USA. Two irrigation misters (Section \ref{sec:filtering}) are fixed at interior positions indicated by the arrows in Fig.~\ref{fig:wall} to simulate leaks. These water lines can be operated independently at utility pressure. The photograph shows the structure with the rear water-repellent gypsum board (thickness: 1.3 cm) attached but prior to sealing it with the second, outer section of gypsum board.

Testing reveals that a single simulated leak can be detected \emph{anywhere} on the span of the wall, despite the presence of up to five vertical studs in the path. The left-most leak in Fig.~\ref{fig:wall} can be detected at the far-right of the structure, but the sensor must be positioned on or very close to the gypsum board. At this distance from the leak, moving the sensor more than a few cm from the wall surface causes loss of the leak signal. We believe the structure is acting as an acoustic chamber and/or waveguide. An evanescent wave couples enough energy across the wall boundary that it can be detected -- even with outdoor background noise present (wind, foliage, birds, insects, street traffic, etc). No access hole in the gypsum board is required to hear leaks behind it.
 
The Helmholtz resonator and microphone assembly (Section \ref{sec:filtering}) introduces some sound collection directivity. A sensor mounted on a wall to monitor for interior leaks could be oriented in a direction opposite room noise to improve reliability. Using multiple sensors for indoor free-space leak triangulation can be problematic, however, due to reverberation, i.e.~sound reflection from surfaces.

\section{Conclusion}

An automated acoustic leak sensor can be a robust solution for difficult applications, especially for alerting to the onset of very small water leaks occurring on pressurized plumbing. Standoff detection allows for versatile placement and wide area coverage; no contact with pipes is necessary or desirable.  Personal security is assured because the device is unable acquire intelligible conversations:~eavesdropping is physically impossible. Careful attention to power management along with extremely low data rates make it a good candidate for deployment as a remote node on a wireless sensor network.

\section{Acknowledgments}

The author thanks Vasudevan Nampoothiri for helpful discussions and Gary Rex for constructing the test structure.


\begin{thebibliography}{00}

\bibitem{1} 
``Low-Power Flood and Freeze Detector Reference Design With Sub-1 GHz and 10-Year Coin Cell Battery Life'', Texas Instruments Design TIDA-01518, 2017 [Online]. Available: \url{https://www.ti.com/tool/TIDA-01518}

\bibitem{2} Li, X.J. Li and P.H.J. Chong, ``Design and Implementation of a Self-Powered Smart Water Meter'', \emph{Sensors} vol. 19, pp. 4177, 2019. [Online]. Available: \url{https://doi.org/10.3390/s19194177}

\bibitem{3} Engineering Acoustics/Outdoor Sound Propagation, Wikibooks [Online]. 
Available: \url{https://en.wikibooks.org/wiki/Engineering\_Acoustics/}

\bibitem{4} T.G. Forrest, ``From Sender to Receiver: Propagation and Environmental
Effects on Acoustic Signals,'' \emph{Amer. Zool.}, vol. 34, pp. 644--654, 1994.

\bibitem{5} Osama Hunaidi, Alex Wang, Marc Bracken, Tony Gambino, and Charlie Fricke, ``Acoustic Methods for Locating Leaks in
Municipal Water Pipe Networks,'' in \emph{International Conference on Water Demand Management,} Dead Sea, Jordan, 2004, pp. 1--14.

\bibitem{6} Fabrício Almeida, Michael Brennan, Phillip Joseph, Stuart Whitfield, Simon Dray, and Amarildo Paschoalini, ``On the Acoustic Filtering of the Pipe and Sensor in a Buried Plastic Water Pipe and its Effect on Leak Detection: An Experimental Investigation,'' \emph{Sensors} vol. 14, pp. 5595--5610, 2014. [Online]. Available: \url{https://doi.org/10.3390/s140305595}

\bibitem{7} Bhargavi Nisarga and Kripasagar Venkat, ``A Robust Glass-Breakage Detector Using the MSP430,'' Texas Instruments Application Report SLAA389, February 2008 [Online]. Available: \url{https://www.ti.com/lit/an/slaa389/slaa389.pdf}

\bibitem{8} MicroPhonon. Acoustic Leak Detector [Online]. 
Available: \url{https://github.com/microphonon/acoustic-leak-detector}

\bibitem{9} Ronald L. Panton and John M. Miller, ``Resonant frequencies of cylindrical Helmholtz resonators,'' \emph{J. Acous. Soc. Amer.}, vol. 57, pp. 1533--1535, 1975.

\end{thebibliography}
\end{document}